# A NEW FRAMEWORK FOR SOFTWARE LIBRARY INVESTMENT METRICS

# A NEW FRAMEWORK FOR SOFTWARE LIBRARY INVESTMENT METRICS

By

**Anas Mohammad Hassan Shatnawi**

**Advisor Name**

**Dr. Ismail I. Hmeidi**

**Co.Advisor Name**

**Dr. Mohammad Q. Shatnawi**

Thesis submitted in partial fulfillment of the requirements for the degree of M.Sc. in Computer Science

At
The Faculty of Graduate Studies
Jordan University of Science and Technology

# A NEW FRAMEWORK FOR SOFTWARE LIBRARY INVESTMENT METRICS

By

**Anas Mohammad Hassan Shatnawi**

<u>Signature of Author</u> ......................................

| <u>Committee Member</u> | Signature and Date |
|---|---|
| Dr. Ismail Hmeidi (Chairman) | ........................................ |
| Dr. Mohammad Shatnawi (Co.Advisor) | ........................................ |
| Dr. Qusai Abuein (Member) | ........................................ |
| Dr. Izzat Alsmadi (External Examiner) | ........................................ |

Dec, 2011

# DEDICATIONS

***This thesis is dedicated to my father, mother, brothers, sisters, and all of my friends, who offered me unconditional love and support throughout the course of this thesis. It is also dedicated to Dr Saed Abd Alazeez who deceased before the end of this thesis.***

# ACKNOWLEDGEMENTS

This thesis could not have been written without Dr. Mohammad Shatnawi who not only served as my supervisor but also encouraged and challenged me throughout my academic program. He and the other committee members, Dr. Ismail Hmeidi, Dr. Qusai Abuein, and Dr Izzat Alsmadi, guided me through the thesis process, never accepting less than my best efforts. I thank them all. Many thanks to Dr Saed Abd Alazeez who helps me, for I have created my idea.

TABLE OF CONTENTS

## LIST OF FIGURES

LIST OF TABLES

# ABSTRACT

# A NEW FRAMEWORK FOR SOFTWARE LIBRARY INVESTMENT METRICS

By

**Anas Mohammad Hassan Shatnawi**

Software quality is considered as one of the most important challenges in software engineering. It has many dimensions which differ from users' point of view that depend on their requirements. Therefore, those dimensions lead to difficulty in measuring and defining the software quality properly. Software quality measurement is the main core of the software quality. Thus, it is necessary to study and develop the software measurements to meet the better quality.

The use of libraries increases software quality more than that of using generic programming because these libraries are prepared and tested in advance. In addition, these libraries reduce the effort that is spent in designing, testing, and maintenance processes.

In this research, we presented a new model to calculate the saved effort that results from using libraries instead of generic programming in the coding, testing, and productivity processes. The proposed model consists of three metrics that are Library Investment Ratio, Library Investment Level, and Program Simplicity. An empirical analyzes has been applied into ten projects to compare the results of the model with Reuse Percent. The results show that the model has better indication of the improvement of software quality and productivity rather than Reuse Percent.

# Chapter One: Introduction

## 1. Introduction

Nowadays, almost every aspect and tale of human lives depend on information and communication technology in vast manner, so the quality of software must be improved to get a better life [1] [28]. Software engineering is a science that manages the software development process to meet high quality with lowest cost [1] [4]. Therefore, the software quality becomes a critical challenge in the software engineering. The researchers have to study and develop new software metrics to meet better software quality [3].

## 2. Software Quality Measurements

Software quality is one of the most important topics in software engineering; it is considered as the main core of competition in the software market [1]. It has many dimensions which differ from user requirements point of view; this leads to many definitions. These definitions can be simplified as Garvin and Juran to Fitness to use or Conformance on Requirements [4]. It is difficult to measure how much such a system is fitness to use. Therefore the researchers used software quality attributes or aspects to measure how much a system fitness to use. For example understandability, traceability, complexity, testability, and so on can be used as software attributes. Software quality is very important for both costumers and manufactures. Costumers need reliable system that easy to learn and use, and manufactures need a reusable system that is easy to maintenance and test [1] [3].

Software quality measurements and metrics are the main core of the software quality; it is autism views and reduces the difference [1]. It was necessary to study and develop the software metrics to meet the good quality. Software measurement is defined as a process of driving the quantities from features of software entity. Software metric is a quantitative indicator of the software and the software production process attributes that is made us able to sense these attributes [3] [21]. Software metrics can be used to measure product such as source code, development process such as design process, and resources such as production cost [3]. There are two problems that are related to any measurement system, which are representation problem and uniqueness problem. The representation problem is a problem that occurs during the process of formatting particular empirical system to numerical system that made us able to sense the attribute of this system. The uniqueness problem is a problem of finding a good scale system that is used to convert the result of such a metric to another representation [1]. Many software metrics that have correlation between each others can be integrated to build a software model [23]. Several kinds of software metrics and models were proposed to measure different types of software quality attributes. For example, Halstead Complexity Model was presented as a complexity measurement [2], and Reuse Level to measure the amount of reuse [10].

## 3. Software Reuse

Software reuse is a process of reusing existing software artifacts during software life cycle. When a systematic reuse is applied the software quality and productivity are improved by reducing the development time, and the cost [5] [22] [24] [25] [27]. Software reuse does not mean that reusing the source code only, but it can be applied to any development phase such as requirement phase by reusing the experiences and documents [23] [24]. This research is interested in software source code reuse. Reusability of software artifacts is a degree of how much such an artifact is suitable to reuse to get the expected benefits [7] [12] [23] [25]. Organizations have to measure the process of software reuse to find the benefits of reusing, which can be done through software reuse metrics [23]. Thus, software reuse metrics became very important research field in the software engineering science [26].

Software reuse and reusability metrics are very important topics in software quality science, which is used to assess the reuse process [1] [26]. Software reuse metric is a quantitative indicator that is used to measure the amount of reuse in the software [9] [26]. Software reusability metric is an indicator that finds the ability of software component or software artifacts for reusing in other system [19] [27].

There are several proposed software reuse and reusability metrics and models. These models can be classified as [23] to amount of reuse metrics, reusability assessment, cost benefit analysis, maturity assessment, and reuse library metrics. Amount of reuse metrics are used to measure the reuse percentage in the software such as [10]. Reusability assessment is used to measure the reusability of software artifacts [19] such as [12]. Cost benefit analysis interested to find the quality and productivity revenue of reuse [23] such as [9]. Maturity assessment used to evaluate the reuse process to improve its weaknesses [26]. Reuse library metrics are used to find the investment of reusing library [23].

In this research the focus will be on Halstead Complexity Model HCM [2] to measure the saved effort that is resulted from library reuse. Halstead Complexity Model divides the text of program code using lexical analyzer into tokens. These tokens are classified into two factors operands and operators. After that, statistical analysis tools are applied on these factors to compute some metrics. Consequently, number of vocabulary, program length, program volume, program level, program difficulty and program effort can be calculated. All of these are related to program complexity [2].

## 4. Motivation

Due to the vast growth of using the library instead of generic programming, the need to study and develop new software metrics is required [28]. The use of library improves software quality and productivity, and decreases software development cost and time. The purpose of this research is to develop and implement a new framework that calculates the saved efforts when libraries are reused. The model is built based on complexity and testability software quality attributes by calculating the complexity and testability cost for the reused library.

## 5. Problem Statement

As it has been mentioned above, the software metrics are very important to asses the software product and the software production process. After the emergence and the frequent use of libraries, the software quality has increased because these libraries are prepared and tested in advance. In addition, these libraries reduce the effort that is spent in coding, testing, and maintenance processes.

Accordingly, there is a need to define a new metric that measure the saved effort by using libraries and the reduction in coding, testing, and maintenance effort. The effort of the testing and maintenance processes takes 50% of the total software cost [8].

## 6. Problem Solution

The model is supposed to define new metrics to measure the saved effort of using the source code metrics. In this research, the focus will be on Halstead model [2] as the main evaluator to measure the library investment.

The model in this context will take the program source code and then apply the Halstead model only on the program source code to get the initial result. After that, the model will take the used libraries and apply the Halstead model on the program source code with classes and methods that are used from the libraries to get the second result. Eventually, the two results will be compared along with additional computations to get the saved effort.

## 7. Research Assumption

In this research, the work has been done along with certain assumptions. For example, the library is prepared and tested in advance, so the developers do not need any effort to test the reused library. Therefore any use of library will reduce testing and development cost.

## 8. Thesis Outline

This thesis is organized as follow. Chapter one includes the introduction of our research and important definitions. In chapter two the literature reviews are discussed, which classified into four types; the complexity metrics, the amount of reuse metrics, the indusial and empirical analyses of software reuse, and reusability metrics. Chapter three shows the model framework structure. Experimental results are discussed in chapter four. And the conclusion and future works are placed in chapter five.

# Chapter Two: Literature Reviews

Software Engineering has gain special attention during the past decade. Certain aspects of software application and related aims have been discussed. A few of these related subjects will be discussed and the focus will be on Software Complexity metrics, the amount of reuse metrics, industrial and empirical analyses of the impacts of software, and Reusability metrics.

## 1. Complexity Metrics

Software complexity metrics are used to measure the software quality based on software source code. In this section, Halstead Complexity Model [2], and Cyclomatic Complexity metric [14] are discussed.

Halstead [2] presented a Software Science concept by introducing Halstead Complexity Model (HCM), which is used to measure the software complexity. Many researchers studied the HCM and they proved that the HMC is closely related to software complexity.

HCM uses a software source code to derive many metrics. HCM divides the text of software source code using lexical analyzer into tokens; these tokens are classified into two factors, operands and operators. Operators include programming language keywords, mathematical and logical operators, and system APIs such as (int, =, class, &). Operands include identifiers, numbers, punctuation, and string literal such as (class name, "Hello", {, 10, 0xfc).

After that, a few statistical analysis tools are applied on these factors to compute several metrics, which are Number of Vocabulary, Program Length, Program Volume, Potential Program Volume, Program Level, Program Difficulty and Program Effort. All of them are related to software complexity [2]. HCM metrics are calculated based on number of operands and operators in the software source code. The following formulas are used to find the HCM metrics:

$VOC = n1 + n2$.

$Len = n1 \log (n1) + n2 \log (n2)$.

$V = (N1 + N2) \log (n1 + n2)$.

$V^* = (2 + n^*) \log (2+n^*)$.

$L = V^*/V$.

$D = V/V^*$.

$E = V/L$.

Where:

n1: numbers of unique operators in the software source code.

n2: numbers of unique operands in the software source code.

N1: the total number of operators with reputation in the software source code.

N2: the total number of operands with reputation in the software source code.

n*: total number of input and output parameters.

VOC: number of vocabulary.

Len: program length.

V: Program Volume.

V*: Potential Program Volume.

L: Program Level.

D: Program Difficulty.

E: Program Effort.

HCM is considered as an easy to calculate measurement; it does not need to depth to logical structure of the software, and does not rely on the type of programming language.

To provide the readers with how the HCM works. The following C++ sample code is used to provide an example below:

```
#include <iostream.h>
void main () {
int X;
cin >> X;
cout<< X+5;
}
```

Table 1: Analyzes of a C++ Sample Code.

| Operators | Operators Frequency | Operands | Operands Frequency |
|---|---|---|---|
| # | 1 | <iostream.h> | 1 |
| include | 1 | main | 1 |
| void | 1 | X | 3 |
| () | 1 | 5 | 1 |
| {} | 1 | | |
| int | 1 | | |
| ; | 3 | | |
| cin | 1 | | |
| >> | 1 | | |
| cout | 1 | | |
| << | 1 | | |
| + | 1 | | |

From table 1, the number of unique operators (n1) is 12 as appeared, the number of unique operands (n2) is 4, the total number of operators (N1) is 14, the total number of operands (N2) is 6, and the total number of input and output parameter (n*) is 1 (X in example). So, the results of HCM are:

VOC = n1 + n2 = 12 + 4 = 16.

Len = n1 log (n1) + n2 log (n2) = 12 log (12) + 4 log (4) = 15.35.

V = (N1 + N2) log (n1 + n2) = (14 + 6) log (12 + 4) = 24.08.

V* = (2 + n*) log (2+n*) = (2 + 1) log (2 + 1) = 1.43.

L = V*/V = 16.83.

D = V/V* = 0.039.

E =V/L = 1.43.

In [14], MaCabe presented a new complexity metric called Cyclomatic Complexity metric (CC), which is considered one of the most important testability metrics and it is used to measure the software complexity, maintainability, and understandability. The CC has a strong correlation with defects rate as many researched considered.

The Cyclomatic Complexity is derived from graph-theorem; MaCabe defined Cyclomatic Complexity as the number of all possible linear paths in a program control graph. The Cyclomatic Complexity is calculated based on conditional of statements. Such as if, for, while and switch cases that led to many branches and are considered as test cases. The following formula is used to calculate the Cyclomatic Complexity:

CC = e – n + 2 * p.

Where:

CC: Cyclomatic Complexity.

e: numbed of edges.

n: number of nodes.

p: number on connected components.

For example, figure 1 shows a program control graph G with vertices V = {A,B,C,D,E,F,G,H} and edges E = {(A,B), (A,F), (B,C), (B,D), (B,E), (C,E), (C,H), (D,E), (D,G), (E,F), (E,G), (F,H), (F,E), (G,H), (G,G)}, where V represents a set of control statements, and E represents a set of linear flow. A graph G starts from a node A and ends at node H.

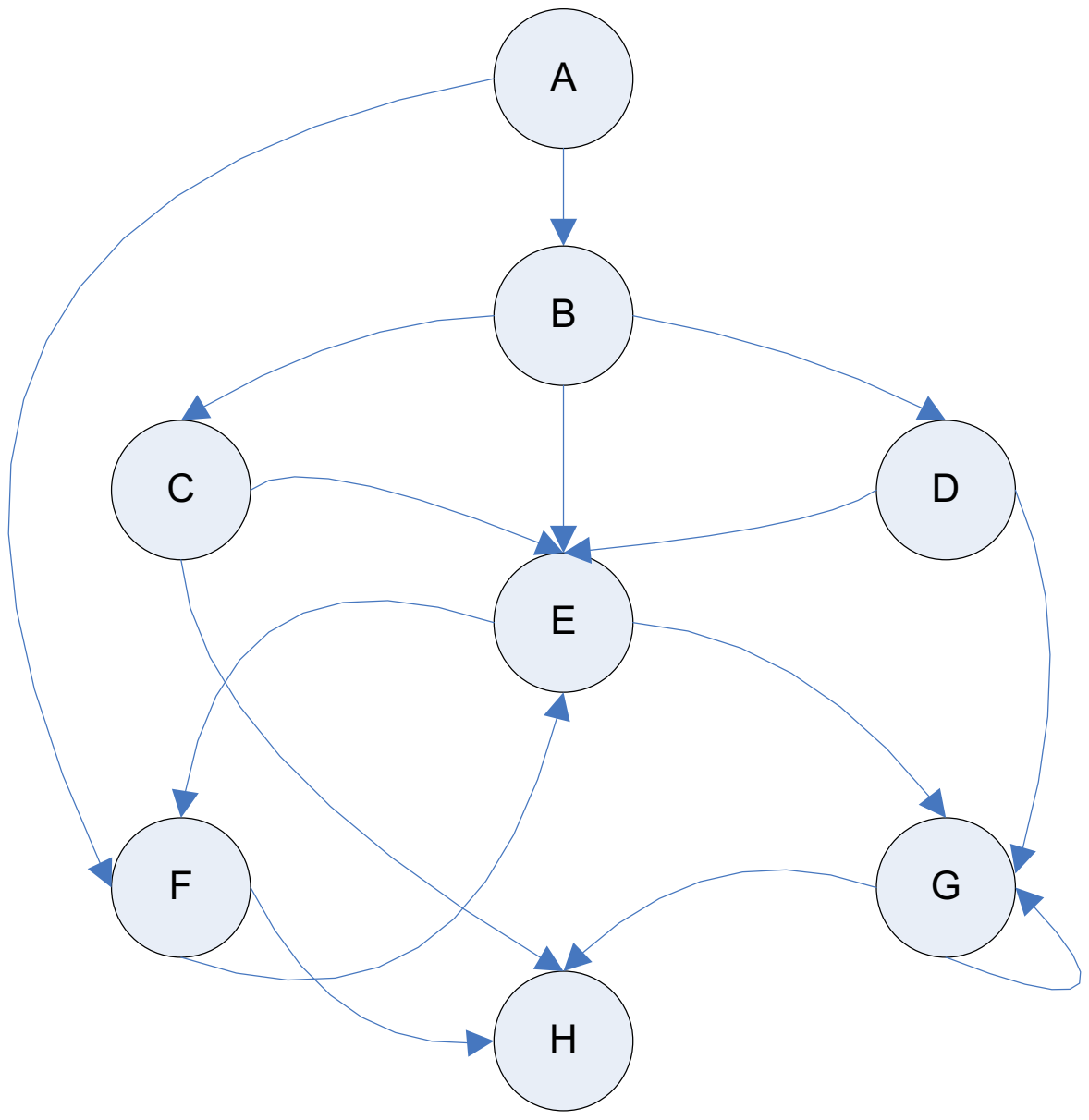

Figure 1: Sample Program Control Graph G.

The parameters of Cyclomatic Complexity equation can be devein from the above program control graph, e = 15, n = 8 and p = 1. Thus, the Cyclomatic Complexity for the above example is:

$$CC = 15 - 8 + 2 * 1 = 9.$$

MaCabe has introduced an easy way to implement and compute the Cyclomatic Complexity based on the number of control statements by finding the numbers of decision points in the program. Decision points are derived from the control statements (e.g. if, if else, while, for, and switch cases). Therefore, Cyclomatic Complexity is n+1, where n is the number of decision points in the program. The CC for the above example can be computed as 8 + 1 = 9.

MaCabe recommends that CC for any program component must be less than 10 to be considered as a good component for testability and maintainability.

## 2. Amount of Reuse Metrics

In this section, the amounts of reuse metrics are Reuse Ratio [6], Reuse Percent [9], Reuse Level [10], and Reuse Frequency [10]. At the end of this section, an example will be provided to explain these metrics.

In [6], a new method of software reuse measurement is presented to predict the saved maintenance efforts, which is the Reuse Ratio. Reuse Ratio is used to asses the effort and cost that are saved when a new software version is created from an existing one to measure the saved testing and maintenance efforts. The aim of this study is to build the reuse based cost model to find the impacts of reuse on quality of the over all software life cycle.

The authors in [6] focus on two factor, files and functions to find the reuse ratio, and divided the files and functions (components) into four types based on reuse ratio. These types are transported, adapted, converted and new one.

When the code of a new component is the same as corresponding component in a previous system then the component is transported component. If the code satisfies reuse ratio from 75% and up to 100%, then, the component is adapted. If it is 50% and up to 75% then, it is a converted component, otherwise the component is a new one and it is considered to be developed from scratch.

Each class has its impacts on system cost (e.g. any component that is considered as transported, there is no need to test it); based on this classification, the authors estimate the saved effort that is produced from software reuse [6].

In [6], authors used SPA tool to find the reuse ratio on TPOCC software in each release from 1.0 to 9.0. SPA provides information about functions and how much they are reused. The same process is applied in each release and next release by finding the amount of reuse using SPA tool. Each function form the target release is compared with the same name function in the last release; if the two functions have the same size (number of line of codes) then, it is considered transported function. If the function has more number of line of code in the later version it assumed to be an adapted function. The function is considered an adapted function if it has smaller size. The new named functions are assumed to be new created functions [6].

In [9], one of the simplest reuse metrics is defined, which is A Reuse Metrics and Return on Investment Model (ROI). ROI model is built based on three metrics, which are Reuse Percent (RP), Reuse Cost Avoidance (RCA), and Reuse Value Added (RVA).

ROI uses some statistical data observations that can be extracted form the software source code. ROI also uses some historical data such as error rate and error cost to apply its metrics, which are:

- Shipped Source Instructions (SSI): refers to the numbers of lines of product code.
- Reused Source Instructions (RSI): refers to the total lines of code for unmodified reused components.
- New and Changed Source Instructions (CSI): refers to the number of new or changed lines in the new releases.
- Software Development Error Rate (TVUA rate): refers to the estimated maintenance cost avoidance in a historical average.
- Software Error Repair Cost (Cost per TVUA): refers to the estimated repair cost in a historical average.
- Source Instructions Reused By Others (SIRBO): refers to the numbers of lines that reused from product in other products.

Reuse Percent (RP) is used as indicator for the reuse amount in the first product, product release, and organization. RP refers to the ratio between the total number of lines of code in the software and the number of reused lines of code. The following equations are used to calculate the reuse percent for the first product, product release and organization respectively:

RP of product = RSI / (RSI + SSI).

RP of product release = RSI / (RSI + CSI).

RP of organization = RSI / (RSI + SSI).

Where:

RP: Reuse Percent.

RSI: Reused Source Instructions.

SSI: Shipped Source Instructions.

CSI: New and Changed Source Instructions.

Reuse Cost Avoidance (RCA) is used to find the financial benefits of reuse by finding the return investment of reuse. The software reuse component needs less cost than creating new software component, but it is not for free. The studies show that its cost and effort is 20% of the total cost of creating new software component. So that, the Development Cost Avoidance can be calculated by the equation below:

DCA = RSI (1 – 0.2) * (CNC).

Where:

DCA: Development Cost Avoidance.

RSI: Reused Source Instructions.

CNC: Cost of New Code.

Maintenance Cost Avoidance is used to find the avoidance cost in software maintenance. MCA is calculated using the following equation:

MCA = RSI * (TVUA Rate) * (Cost per TVUA).

Where:

MCA: Maintenance Cost Avoidance.

RSI: Reused Source Instructions.

TVUA Rate: software development error rate.

Cost per TVUA: software error repair cost.

Then, the RCA is:

RCA = DCA + MCA.

Where:

RCA: Reuse Cost Avoidance.

DCA: Development Cost Avoidance.

MCA: Maintenance Cost Avoidance.

Reuse Value Added (RVA) is used to measure how much the organization uses software reuse. RVA is calculated by the equation below:

RVA = ((SSI + RSI) + SIRBO) / SSI.

Where:

RVA: Reuse Value Added.

SSI: Shipped Source Instructions.

RSI: Reused Source Instructions.

SIRBO: Source Instructions Reused by others.

Reuse Level (RL) and Reuse Frequency (RF) were proposed by authors in [10]. It is assumed that the system is a combination of components at different abstraction level (e.g. a system is combination of functions and each function is a combination of line of code). In this metric, the functions are used as abstraction level. After that, the reused components are divided into two classes, which are the internal and external components. The internal component is a new procedure that is written for current system but it is called several times and the external component is used from other systems or repositories.

The authors in [10] defined data observations that are used to calculate the reuse level, which are internal threshold level (ITL), external threshold level (ETL), number of internal components (IU), number of external components (EU) and total number of components (T). ILT is the minimum number of usage of a specific internal component to be considered as reused component. ETL is the minimum number of usage of specific external component to be considered as reused component. IU is the number of internal component that satisfy ITL. EU is the number of external components that satisfy ETL. T is the total number of components in the system.

Now the reuse level is calculated using the following equations:

IRL = IU/T.

ERL = EU/T.

TRL = (IU + EU)/T.

Where:

IRL: Internal Reuse Level.

ERL: External Reuse Level.

TRL: Total Reuse level.

IU: number of internal components.

EU: number of external components.

T: total number of components.

Based on the above equations, the Reuse Level will be between 0 and 1, when RL = 0, then there is no reuse.

Reuse Frequency (RF) is used to measure the component references by counting the number of used for each components. The authors in [10] defined three variables that are used to measure RF. These variable are number of used of internal component (IUF), number of references of external component (EUF), and total number of references for internal and external (TF). The formula of reuse frequency, internal, external and total reuse frequency are:

IRF = IUF/TF.

ERF = EUF/TF.

TRF = (IUF + EUF)/TF.

Where:

IRF: Internal Reuse Frequency.

ERF: External Reuse Frequency.

TRF: Total Reuse Frequency.

IUF: number of used of internal component.

EUF: number of used of external component.

TF: total number of used for internal and external components.

Now, an example for the above amount of reuse metrics is shown below. The focusing is on the Reuse Percent (RP), Reuse Level (RL), Reuse Frequency (RF), and Reuse Density (RD). Figure 2 shows the structure of the sample software, which contains eight components. Five of them as new components (A, B, C, D, and E) and the other three as reused components (F, G, and H), where each box refers to a component that is described using three parameters e.g. name, size, and status. The status of a component can be a reused components or a new created component. The arrows are referred to the components reused references, and the size of the components is measured using number of lines of code (LOC).

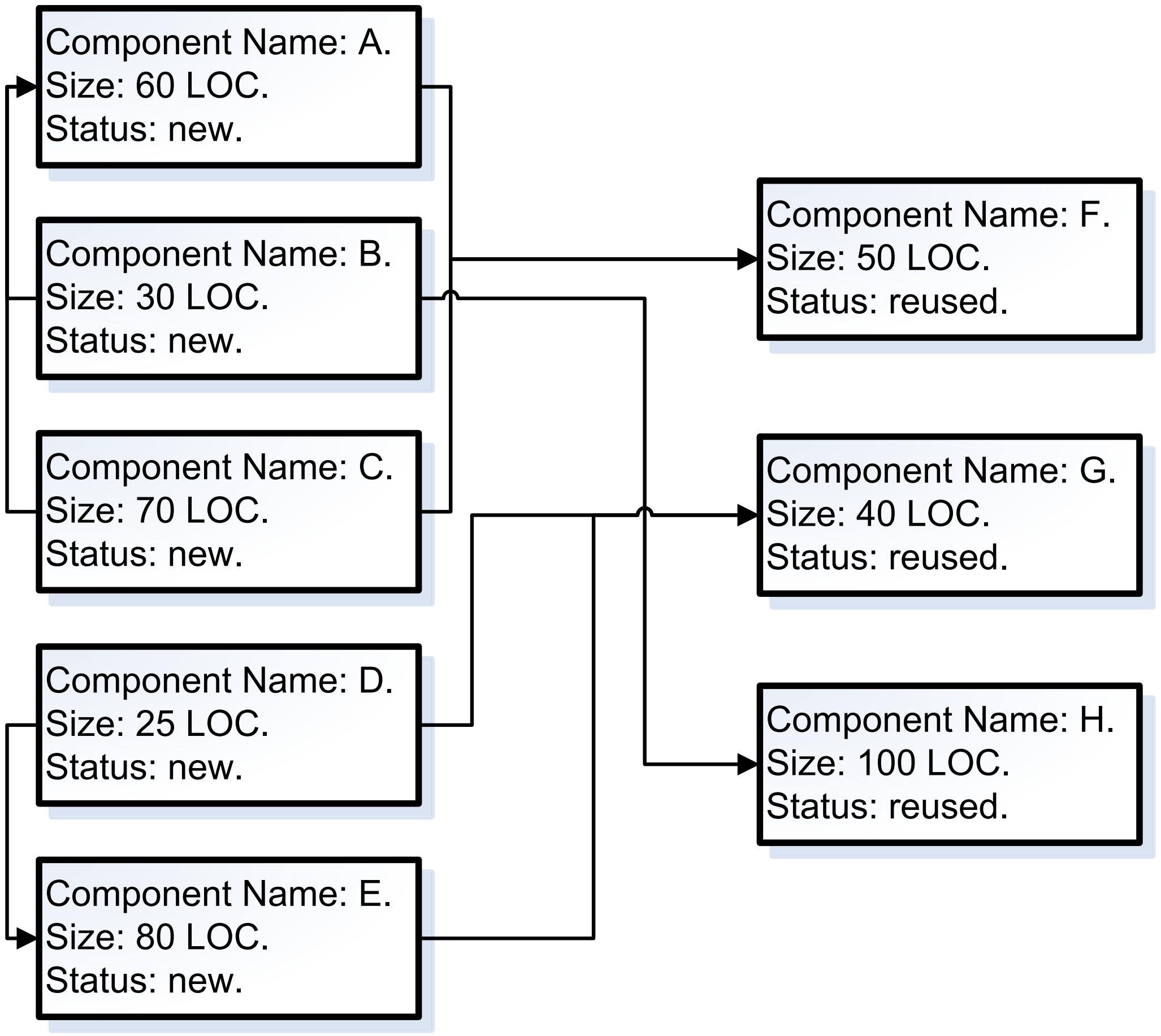

Figure 2: Sample Software Structure.

The Reuse Level (RL) for the sample software is calculated as appears below after assuming that the internal threshold level is 1 and external threshold level is 1:

Number of internal reused components (IU) = 2 components.

Number of External Reused Components (EU) = 3 components.

Total Number of Reused Components (T) = 8 components.

Internal Reuse Level = IU / T = 2 / 8 = 0.25.

External Reuse Level = EU / T = 3 / 8 = 0.375.

Total Reuse Level = (IU + EU) / T = (2+3) / 8 = 0.625.

The Reuse Frequency (RF) is calculated as follow:

Number of References of Internal Reused Component (IUF) = 3 references.

Number of References of External Reused Component (EUF) = 3 references.

Total Used for Internal and External (TF) = 8 references.

Internal Reuse Frequency = 3 / 8 = 0.375.

External Reuse Frequency = 3 / 8 = 0. 375.

The Total Reuse Frequency = (IUF + EUF) / TF = (3 + 3) / 8 = 0.75.

The size of reused component and the size of new created components need to be calculated to find the Reuse Percent (RP), RP for product, product release, and organization are calculated as follow:

Sum of Shipped Source Instructions (SSI) = 265 LOC.

Sum of Reused Source Instructions (RSI) = 190 LOC.

Reuse Percent = RSI / (RSI + SSI) = 190 / (190 + 256) = 0.41.

The Reuse Density RD is calculated as follow:

Number of Internal Reused Components = 2 components.

Number of External Reused Components = 3 components.

Total Number of LOC of all Components= 455.

Internal Reuse Density = 2 / 455 = 0.0043.

External Reuse Density = 3 / 455 = 0.0065.

The Total Reuse Density = Internal Reuse Density + External Reuse Density = 0.0043 + 0.0065 = 0.0108.

## 3. Industrial and Empirical Studies of Software Reuse Impacts

Five industrial and empirical analyses of the impacts of software reuse are discussed in this section.

A study and analysis of amount of reuse metrics have been applied in [11] to extract relationships between reuse metrics and other software complexity metric like Cyclomatic

Complexity and Line of Code. The experimental data consist of 70 projects that are collected from public libraries. The author in [11] selected some reuse metrics, which are:

- Reuse Level: refers to the ratio between the numbers of reuse components to the number of total components [8] [11].
- Reuse Frequency: refers to the ratio between the numbers of references of reused components to the total number of references in the system [8] [11].
- Reuse Density: used to measure the reuse density in the system by finding how much the reuse is used [8] [11].

The results show that there is a little relation between Reuse Level (RL) and the number of lines of code (LOC). Reuse Frequency (RF) has also weak relation with LOC, but there is a strong correlation between Reuse Density and LOC. Reuse Level, Reuse Frequency, and Reuse Density have different correlation between each other. The relationship between RL and RF is high average in this case (0.97). The correlation between RL – RD and RF – RD is a medium correlation in this case (0.47) and (0.56) respectively [11].

The authors in [15] presented the reuse and its impacts on software quality and productivity by analyzing four data sets of C and C++ components that gathered from industrial companies.

In [15], several data variables are used to measure the software quality and productivity, which are:

- Number of line of code (NCSL) to measure the component size by finding the number of semicolons in a components source code file.
- Error density that refers to number of errors per NCSL.
- Number of component deltas. Delta refers to a change in a component, where change may lead to a fault.

- Number of NCSLs that are produced by a person per day. It is used to measure the saved effort.
- Quality perceived by developers. It depends on developer's experience in software maintaining and debugging.

In the study in [10], the authors use the Reuse Level (RL) and the Reuse Frequency (RF) as main amount of reuse measurement. The relationship can be determined between software quality and productivity and the amount of reuse using Spearman's rank correlation [16]. The result shows that the software quality is increased with software reuse. More reuse means more quality, but there is an ambiguity in the relationship between productivity and amount of reuse.

In [17], a new development process was proposed, in which it can manage the reuse process. This development process is called the Reuse Oriented Development model (ROD). ROD manages the reuse process by storing, retrieving, and searching software artifacts in a repository. These artifacts can be reused in a new software. Each artifact has a description, input, and output parameters. The developers search for the needed component in a repository before beginning to develop a new one.

After that, an industrial study was applied on two ongoing projects from Italian Small Medium Enterprise (SME) to investigate the impacts of Reuse Oriented Development model (ROD) on software quality and productivity. One of the selected projects has been worked under ROD, and the second project has been worked under conventional CONV way. Both projects were implemented using COPOL programming language. The experiment was run three times during software development phases. In each run, the authors applied some metrics that divided into two product metrics and one process metrics [17].

Product metric is used to measure product quality by finding mean Cyclomatic Complexity for each component by calculating the summation of Cyclomatic Complexity in the overall system divided into number of components. The results show that the mean Cyclomatic Complexity for ROD is less than that for the CONV; in other words, the quality of ROD project is higher than CONV project [17].

Process metrics is applied to find the reuse impacts on software productivity using Apparent Productivity metric, which measures the amount of produced code for a person per hour. The formula of Apparent Productivity is:

AP = (NCSL + NRLOC)/ ESPH.

Where:

AP: Apparent Productivity.

NCSL: Number of Line of Code.

NRLOC: Number of Reused Lines of Code.

ESPH: Effort Spent by a Person per Hour.

The results show that the ROD has large Apparent Productivity compared to CONV project. Actual Productivity is other Process metric, which measures the amount of produced code for a person per hour without taking the software reused into account. The formula is:

ACP = NWLOC/ESPH.

Where:

ACP: Actual Productivity.

NWLOC: Number of Written Line of Code.

ESPH: Effort Spent by a Person per Hour.

The results show that the ROD productivity in the first two runs is high, but it is decreased in the third run because the developers begin to reuse components from repository. This measures that the ROD quality should be better than that from the CONV over project life cycle [17].

In [18], an analytical and empirical study was applied on many software reuse metrics to evaluate how it measures the impacts of software reuse. This reuse metrics include the Reuse Level (RL), the Reuse frequency (RF), the Reuse Ratio (RR), and the Reuse Source Instruction (RSI). The authors defined a reuse benefits as:

Rb(S) = C(S without reuse) – C (S with reuse)/ C(S without reuse).

Where:

Rb: Reuse Benefits.

C: Development Cost.

S: System.

The authors in [18] introduced some properties that were used to evaluate and analyze the target metrics, which are:

- Property 1: maximum and minimum value of Rb for every system. The value of Rb should be started form 0 to the value less than 1. When Rb is 0, then there are no any benefits. For every system S, 0 >= Rb > 1.
- Property 2: implementation dependency, in which it is possible to have two systems with the same functionality, but with deferent implementation. Thus, they have different Rb.
- Property 3: any system can be implemented in different ways with different reusing based on the style of implementation. Each system S, such that Rb(S)>0, there is another system S' that has the same functionality, and Rb(S) > Rb (S').
- Property 4: it is used to evaluate the sensitivity of reuse benefits to the number of such a component is reused. Thus, when a system is re-implemented by reducing the number of reusing such a component C then the reuse benefits of new system should be less than the old one.

- Property 5: the Rb of a system is sensitive to the cost of reused components. Suppose that, there are two components C1 and C2 that have the same functionality but with different cost, where Cost (C1) > Cost (C2). A system S implemented two times. C1 is used in the first implementation, and C2 is used in the second implementation. Thus, Rb (S with C1) > Rb (S with C2).
- Property 6: reusing of external component is better than reusing internal component, so the Rb of a system S is better when it uses external component.
- Property 7: a system S reuses external component n times with cost C1. The Rb of S should be decreased when the developer replaces one reused time of this component with another external component that has less cost.
- Property 8: it is concerned with cut and paste reused. Suppose that, there are three instances of a system S1, S2, and S3, where S1 is implemented without reusing, S2 is implemented with reusing such a component that in slightly modified, and S3 is implemented with reusing such a component without any modification. Accordingly Rb (S1) < Rb (S2) < Rb (S3).

The analytical result shows that each metric has satisfied some properties, which rely on its weaknesses, but there is no such a metric that apply all properties. All target metrics are conformed to properties 2 and 3. Table 2 summarizes the results of analytical study for all target metrics.

Table 2: the Analytical Results of [18].

| Property | RL | RF | RSI | RR |
|---|---|---|---|---|
| 1 | Partial conformance | Partial conformance | Partial conformance | Partial conformance |
| 2 | Conform | Conform | Conform | Conform |
| 3 | Conform | Conform | Conform | Conform |
| 4 | Partial conformance | Partial conformance | Non conformance | Non conformance |
| 5 | Non conformance | Non conformance | Partial conformance | Partial conformance |
| 6 | Partial conformance | Partial conformance | Conform | Conform |
| 7 | Conform | Conform | Non conformance | Non conformance |
| 8 | Non conformance | Non conformance | Non conformance | Partial conformance |

An empirical study also has been applied in [18] to validate the target metrics experimentally by finding the impacts of the above metrics on the software quality and software productivity. The experimental data was collected from 7 student projects in Maryland University. The productivity is measured by finding the ratio between the system size (Line of Code) and the total time spent during the system development life cycle (Hours), and the quality is measured using rework efforts, errors, and faults density.

The results can be shown by discussing the relationship between reuse metrics in the first hand and quality and productivity in the other hand. Reuse Ratio (RR) has the best correlation with productivity and well relationship with faults and errors density, but little with rework efforts. Reuse Level and Reuse Frequency are correlated to rework efforts, errors and faults density [18].

An empirical case study was presented by Succi in [20] to analyze the impacts of software reuse on customer satisfaction in an RPG environment. Succi was used several software metrics to measure the amount of reuse, software size and complexity, and customer satisfaction for the experimental data. Table 3 shows the used metrics in the proposed system.

Table 3: the Used Metrics in the System Proposed by [20].

| Software Attribute | Metrics |
|---|---|
| Software size and complexity | Line of Code, Program Volume and Cyclomatic Complexity. |
| Amount of reuse | Reuse Level, Reuse Density and Reuse Frequency. |
| Customer satisfaction | Customer Complaint Density. |

Customer Compliant Density measures the customer satisfaction through customer survey or by finding the ratio between numbers of customer complaints per file or system size, but the company should collect these complaints in a list. Customer Compliant Density is calculated using the following formula:

CCD = # of Customer Complaints / # of Line of Code.

The study was applied on two accounting system that are selected from Italian Company, the first system is implemented using ad hoc reuse strategy, and the second one is implemented using library of reusable components strategy.

The results show that there is a positive correlation between amount of reuse metrics and customer satisfaction. Customer satisfaction is increased when the developer uses a library reuse.

## 4. Reusability Metrics

Reusability metric is an indicator that finds the ability of software component for reused [19]. In this section, four reusability metrics are discussed.

In [7], Function Template Factor (FTF) and Class Template Factor (CTF) metrics are produced. These metrics measure the amount of reusability of function and class template. The FTF metric measures the reuse of function template by finding the ratio between the number of functions that use function templates and the total number of function. The CTF metric measures the reuse of class template by finding the ratio between the numbers of classes that use this class template to the total number of classes [7].

In [12], Global Coupling Metric was presented; this metrics measure the relationship between various modules and its data. The relationship occurs between module A and module B if module A uses one or more data or function member in module B. this metric takes into account the direct and indirect relationship, if A depends on B and B depends on C, then A depends on C as well [12].

Components Reusability Metric presented in [13]. This metric considers four factors to establish the reusability of components, which are customization of component, interface complexity, portability and documentation quality. Customization is measured by assess the ability to modify the component as which the application needs. Interface complexity should be well defined, simple, and understandable. Portability is measured by checking if the component can be worked on different platforms. The documentation quality should be easily to understand by the user and then understanding the component features [13].

Gandhi and Bhatia in [19] introduced four new metrics to measure the reusability of templates in object oriented software, which are Number of Template Children (NTC), Depth of Template Tree (DTT), Method Template Inheritance Factor (MTIF), and Attribute Template Inheritance Factor (ATIF).

- Number of Template Children (NTC): for a super template class C, NTC is number of subclasses that have instances of C.
- Depth of Template Tree (DTT): it is the maximum path from a class to a super class in the inheritance graph.
- Method Template Inheritance Factor (MTIF): it is calculated by finding the ratio between the numbers of methods that inherit from super template classes to the numbers of variable methods in all classes.
- Attribute Template Inheritance Factor (ATIF): it is calculated by finding the ratio between the numbers of attributes that inherit from super template classes to the numbers of variable attribute in all classes.

The results show that the applicability of the proposed metrics is just used for template style systems [19].

# Chapter Three: A New Framework for Software Library Investment Metrics

## 1. Introduction

In the previous chapters, the important information that is related to the measurement model has been mentioned. In this chapter, the measurement model that contains three library investment metrics is introduced. The three investment metrics are Library Investment Ratio (LIR), Library Investment Level (LIL), and Program Simplicity (PS). The metrics are discussed at the first section, the structure of model is placed in section two, and an example is presented in section three.

## 2. Library Investment Model

The developed and implemented model contains three library investment metrics, which were derived based on Halstead Complexity Model [2]. Halstead presented the concept of potential program volume as a perfect program volume that is implemented using a typical programming language that can represent the needed operators by predefined functions without any need to implement the algorithms. It needs only to identify the operands. Therefore, the software quality is better whenever the program volume is closely to potential program volume, which can be done through library reuse.

The model presents three metrics, which are Library Investment Ratio (LIR), Library Investment Level (LIL), and Program Simplicity (PS), that are calculated based on Program Volume (V). The following formula is used to find the program volume:

$V = N \log (n).$

Where:

V: Program Volume.

n: is the number of unique operands and operators.

N: is the total number of operands and operators with frequent.

The model depends on three parameters, which are original program volume (Vorg) that comes from library reuse, program volume without library reuses (Vnr), and the reduction volume (Vr) that is resulted from library reuse. These volumes are calculated using the following formulas:

$Vr = \sum (fci * Vci)$.

$Vci = Nci \log (nci)$.

$Vorg = N \log (n)$.

$Vnr = Vorg + Vr$.

Where:

Vr: the reduction volume that resulted from library reuse.

fci: the frequent number of used of library component c.

Vci: the volume of library component c.

i: refers to a series of library components.

Nci: is the total number of operands and operators with repetition in a component c.

nci: is the number of unique operands and operators in a component c.

Vorg: program volume for original program.

n: is the number of unique operands and operators.

N: is the total number of operands and operators with repetition.

Vnr: program volume without reuse.

Now, the following sections presented the developed metrics, starting from the Library Investment Ratio Metrics.

## 2.1 Library Investment Ratio Metric (LIR)

The LIR metric is developed to measure the reduction volume ratio that is resulted from using library instead of generic programming. Generic programming is the programming pattern that has not used any library reuse.

LIR represents the ratio between Vr and Vnr. Vr is the program volume that is resulted from library reuse. Vnr is the expected program volume that is resulted without library reuse (Vnr).

The formula of LIR is:

LIR = Vr / Vnr.

Where:

LIR: Library Investment Ratio.

Vnr: program volume without reuse.

Vr: the reduction volume that resulted from library reuse.

LIR metric is used to measure the reduction in software complexity, software design cost, and software testing cost that are resulted from library reuse. The LIR result range should be between zero and one (0 - 1). The worst case of LLR is 0, where the library reuse has never been used (Vr = 0). The value one is unachievable because it means that there is no a new software and it is used an existing software (Vr = Vnr). Therefore, the LIR value is better whenever it is increased as much as possible.

## 2.2 Library Investment Level Metric (LIL)

Library Investment Level is used as indicator to the investment level that is resulted from library reuse. Investment level refers to the reduction level that resulted from reusing library. LIL is used to measure the improvement in software productivity. LIL is the percentage between Vr and Vorg. Vr is the reduction program volume that is resulted from library reuse. Vorg is the program volume of current program lonely.

LIL is computed using the following equation:

LIL = Vr / Vorg.

Where:

LIL: Library Investment Level.

Vorg: program volume for original program.

Vr: the reduction volume that is resulted from library reuse.

The minimum value of LIL is zero (i.e. when Vr = 0). The zero value means that the library has not been invested. LIL is increased whenever library reusing increases. This metric can be used as a factor that helps the decision maker to manage the available resources to improve its productivity.

## 2.3 Program Simplicity Metric (PS)

Program Simplicity metric is used to measure the simplicity ratio that is resulted from library reuse.

The formula of PS is:

PS = 1 – (Vorg / Vnr).

Where:

PS: Program Simplicity.

Vnr: program volume without reuse.

Vorg: program volume for original program.

PS value should be between zero and one ($0 <= PS < 1$), the value zero means that there is no any simplicity (i.e. where Vorg = Vnr), the higher PS means more simplicity ratio.

## 3. The Model Structure

In this section, the model structure is discussed. The model structure appears in figure 3.

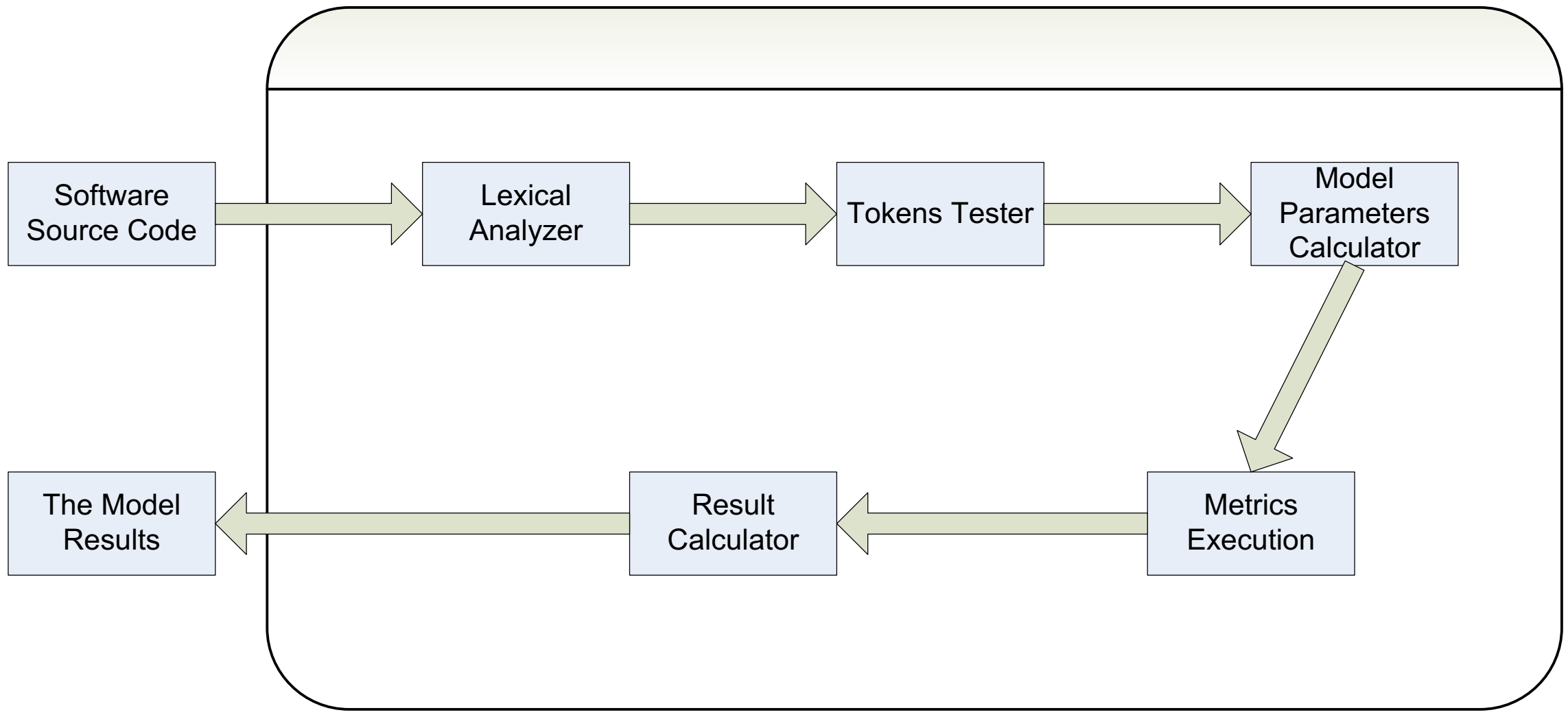

Figure 3: the Model Structure.

The model works as follow. The software source code files are sent to the lexical analyzer that divides the source code into tokens. The Token tester testes the program tokens to classify them into operands, and operators. Model Parameters Calculator calculates Vr, Vorg, and Vnr. After that, the Metrics Execution applies the model equation to find the result of LIR, LIL, and PS. In the last step, the model results are output using Result Calculator.

## 3.1 The Model Main Algorithm

The main algorithm that is used to implement the model is presented in this section. It is used to manage the model components. The algorithm is:

```
Library Investment Model (){
        Tokens = Lexical Analyzer (source code files)
        Oprnds_Oprtrs_List = Token Tester (Tokens)
        Vorg_Vr_Vnr_Vector = Model Parameters Calculator (Oprnds_Oprtrs_List)
        Result = Model Execution (Vorg_Vr_Vnr_Vector)
        Print (Result)
}
```

## 3.2 Lexical Analyzer

Lexical Analyzer is used to scan the target software source code to divide the program source code into tokens. These tokens can be operators, identifiers, keywords, digit, and so on. After that, the comment statements are removed by the lexical analyzer.

The algorithm of lexical analyzer is:

```
Lexical Analyzer (source code files) {
        Ch = getCh ()
        While (Ch != end of file)
        {
                If (is part of (operator))
                {
                        OperatorToken =OperatorToken + Ch
                }
                Else If (is part of (operand))
                 {
                        OperandToken = OperandToken + Ch
                }
                Else If (is separator (Ch)
                {
                        Save current token and establish a new search for a new
                        token
                }
                Ch = get character from input file
        }
}
```

## 3.3 Tokens Tester

Token Tester is used to test the program tokens that are resulted from the lexical analyzer to find the needed factors. These factors are operands, operands frequency, operators, and operators' frequency.

The algorithm of Token Tester is:

```
Token Tester () {
        Cur = Get token ()
        While (Cur != NULL) Do
        {
                If (isOperator (Cur))
                {
                        OperatorsList.Add (Cur)
                        OperatorsCounter = OperatorsCounter + 1
                }
                Else If (isOperand (Cur))
                {
                        OperandsList.Add (Cur)
                        OperandCounter = OperandCounter + 1
                }
                Cur = next token
        }
}
```

## 3.4 Model Parameter's Calculator

Model Parameter's Calculator is used to find the parameters that are used in the introduced metrics, which are, Vr, Vorg, and Vnr. These parameters are calculated based on the results of Token Tester, which are operands, operands frequency, operators, and operators' frequency. It uses the following equations:

$Vr = \sum (fci * Vci)$.

$Vorg = N \log (n)$.

$Vnr = Vorg + Vr$.

Where:

Vr: the reduction volume that is resulted from library reuse.

fci: the frequent number of references of the library component c.

Vci: the volume of library component c.

i: refers to series of library components.

Vorg: program volume for original program.

n: is the number of unique operands and operators.

N: is the total number of frequent of the operands and operators.

Vnr: program volume without reuse.

## 3.5 Metrics Execution

Metrics Execution applies the equations of the metrics to find the model results. It takes Vr, Vorg, and Vnr, which are resulted from the Model Parameter Calculator. The results of the Metrics Execution are the LIR, LIL, and PS. The following formulas are used to find these metrics:

$LIR = Vr / Vnr$.

$LIL = Vr / Vorg$.

$PS = 1 – (Vorg / Vnr)$.

Where:

LIR: Library Investment Ratio.

LIL: Library Investment Level.

PS: Program Simplicity.

Vnr: Program Volume without reuse.

Vorg: Program Volume for original program.

Vr: the reduction volume that is resulted from library reuse.

### 3.6 Result Calculator

The purpose of the Result Calculator is to extract the final results of the metrics. The results include the Library Investment Ratio, Library Investment Level, and Program Simplicity.

## 4. Example of The Introduced Metrics

In this section, an example that describes the process of the model is introduced and discussed. A sample of source code that is written in C++ programming language is used to test the model metrics. In addition to, a comparison benchmark is used to compare the results of the model with Reuse Percent (RP).

Reuse Percent is proposed by [9]. It is used as indicator for the reuse amount in the software source code. RP is the ratio between the total number of lines of code in the software and number of reused line of code. The formula of RP is:

RP = RSI / (RSI + SSI).

Where:

RP: Reuse Percent.

RSI: Reused Source Instructions.

SSI: Shipped Source Instructions.

The sample is presented in table 4. It uses stack file as a library file by calling Stack.h. The lower case of alphabetic "h" refers to the header file in the C++. The sample code creates an object that is belonged to the stack type. After that, the numbers (between 0 and 1) are pushed to the stack object. Then, the program retrieves the stack value using pop method to print the popped results.

Table 4: C++ Sample Program.

| **Original program** | **Stack file that is used from the library** |
|---|---|
| #include <iostream.h><br>#include "Stack.h";<br>int main( )<br>{<br>Stack<int> s;<br>int stackSize;<br>cout<<"Enter Stack Size:";<br>cin>>stackSize;<br>for( int i = 0; i < stackSize; i++ )<br>s.push( i );<br>while( !s.isEmpty( ) )<br>cout << s.Pop( ) << endl;<br>return 0;<br>} | Stack:: Stack()<br>{<br>topOfStack = -1;<br>}<br>bool Stack::isEmpty( )<br>{<br>return topOfStack == -1;<br>}<br>bool Stack:: isFull( )<br>{<br>return topOfStack == SIZE - 1;<br>}<br>void Stack::makeEmpty( )<br>{<br>topOfStack = -1;<br>}<br>int Stack::pop( )<br>{<br>return theArray[topOfStack--];<br>}<br>void Stack::push(int & x ) |
| | {<br>theArray[ ++topOfStack ] = x;<br>} |

Lexical Analyzer parses the source code into tokens; these tokens are the input to the Token Tester. Token Tester testes the tokens and divides them into operands and operators. Table 5 shows the results of Token Tester for both original program source code and the used methods from the library. For the above example the used methods from the stack file are the stack constrictor, pop, push, and isEmpty methods. Therfore, these methods are only considered when calculating the Vr.

Table 5: Token Tester Results.

| | Original Program | The Used Functions |
|---|---|---|
| # of Unique Operands | 7 | 8 |
| Total # of Operands with Frequent | 12 | 15 |
| # of Unique Operators | 17 | 13 |
| Total # of Operators with Frequent | 37 | 18 |

Model Parameter's Calculator uses the results of Token Tester to find the model parameters. The model parameters are Vorg, Vr, and Vnr. Vorg is the program volume of original program source code only. Vr is the reduction volume that is resulted from the library reuse. Vnr is the expected program volume of the software without reusing any library, in which the Vnr = Vorg + Vr. The results of the Model Parameters Calculator are:

Vorg = 49 log (24) = 67.63.

Vr = 33 log (21) = 43.63.

Vnr = Vorg + Vr = 111.26.

Metrics Execution is used to find the values of the model metrics, which are:

LIR = Vr / Vnr = 43.63 / 111.26 = 0.39.

LIL = Vr / Vorg = 43.63 / 67.63 = 0.64.

PS = 1 – (Vorg / Vnr) = 1 – (67.63 / 111.26) = 0.39.

From the above results; the LIR indicates the reduction ratio in software complexity, software design cost, and software testing cost. In this case, 39% is the reduction ratio that is resulted from reusing library. LIL is 0.64, which indicates the improvement level in software productivity. The simplicity that comes from library reuse is 0.39 based on the PS result.

Reuse Percent is calculated by finding two factors. The factors are number of reused source instructions (RSI), and shipped source code (SSI). RSI is the numbers of line of code for methods that are used from the library. SSI is the numbers of line of code for the original program only. In this case, RSI = 4, and SSI = 14. Thus, the RP is:

RP = 4 / 14 = 0.28.

By comparing the results of Library Investment Ratio and Reuse Percent, the conclusion can be drown, in which the LIR indicates that 39% of reuse percentage and 28% for RP. The large gab between them is generated from the differences in the calculation methods. RP finds the reuse ratio based on the numbers of line code, but LIR deepens in the line of code by taking the content of the line of code in its consideration. Therefore, the results of the model metrics are better than Reuse Percent.

# Chapter Four: Experimental and Results

In order to evaluate the model metrics, an empirical and analysis study is applied on several software projects that are collected from Maysalward Inc (MRD) [29]. The Java programming language [30] is used to implement the model. The results of Library Investment Ratio metrics is compared with the results of Reuse Percent (RP), which is proposed by [9].

## 1. Data Collection

Ten projects that developed by Maysalward are collected. MRD is a small size Jordanian company that develops mobile and online games. MRD has about 20 employees.

The gathered projects are belonged to several game categories. These categories are card, puzzle, arcade, sport, and educational one. Each project has been developed by different teams. These teams have different technical and programming skills. Table 6 shows some descriptive statistical information about the gathered projects.

Table 6: Descriptive Statistical Information about Projects.

| Project Name | # of Line of Code | Category |
|---|---|---|
| Arcanoid | 6091 | Arcade |
| Balot | 18458 | Card |
| Carrom | 6973 | Board |
| Fruity | 3906 | Educational |
| Goal Englizi | 7942 | Sport |
| Loteria | 6255 | Card |
| Minesweeper | 2752 | Puzzle |
| Tarneeb | 8546 | Card |
| Taxi Escape | 3637 | Arcade |
| Trix | 14327 | Card |

Holodeck is a library that is developed by MRD. Holodeck is used to develop the gathered projects.

The numbers of Line of Code (LOC) for experimental projects is shown in figure 4. The blue line refers to the numbers of LOC for original programs only. The red line refers to the numbers of reusing LOC from the library.

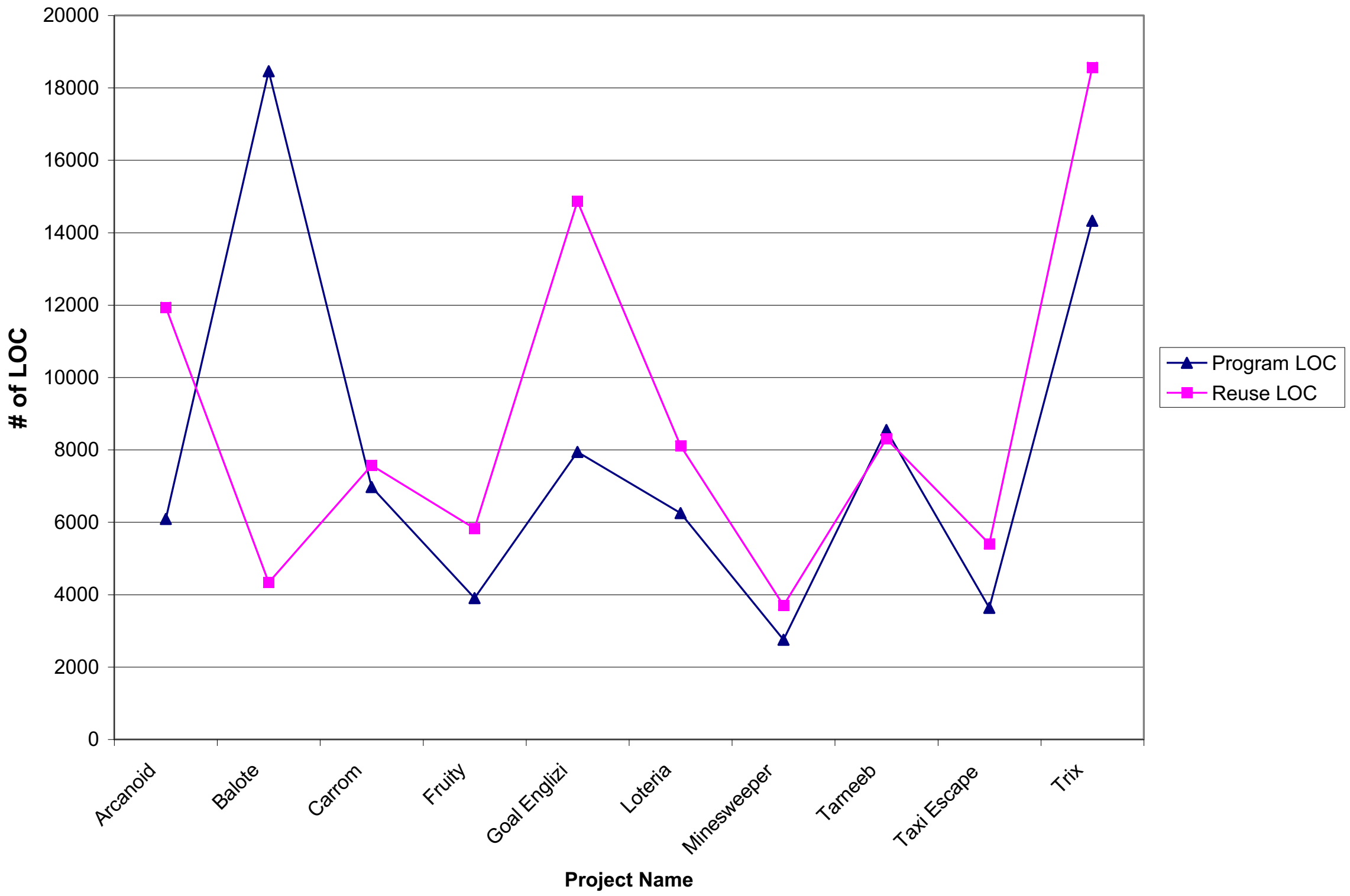

Figure 4: the Numbers of LOC for Original Program and Numbers of Reusing LOC.

## 2. The Implementation of The Evaluation Tool

The evaluation tool is implemented using Java programming language. It is used to evaluate the model metrics and to compare the model with Reuse Percent.

The aim of this tool is to find the results of model metrics. It takes two input parameters that are project source code, and library source code. The output is the values of model metrics, which are Library Investment Ratio (LIR), Library Investment Level (LIL), and Program Simplicity (PS).

## 3. Experimental Results

The results of applying the model metrics into the target projects are presented and discussed in this section.

Figure 5 shows the program volumes for the experimental projects. These volumes are Vorg, Vr, and Vnr. Vorg is the program volume of the original source code of the project only. Vr is the program volume of the source code of the methods that are used from the library. Vnr is the expected program volume of the project without reusing any library. The results show that Trix has the highest reduction volume Vr, and then Goal Englizi, Arcanoid, Tarneeb, Loteria, Carrom, Balot, Fruity, Taxi Escape, and Minesweeper in decreasing order.

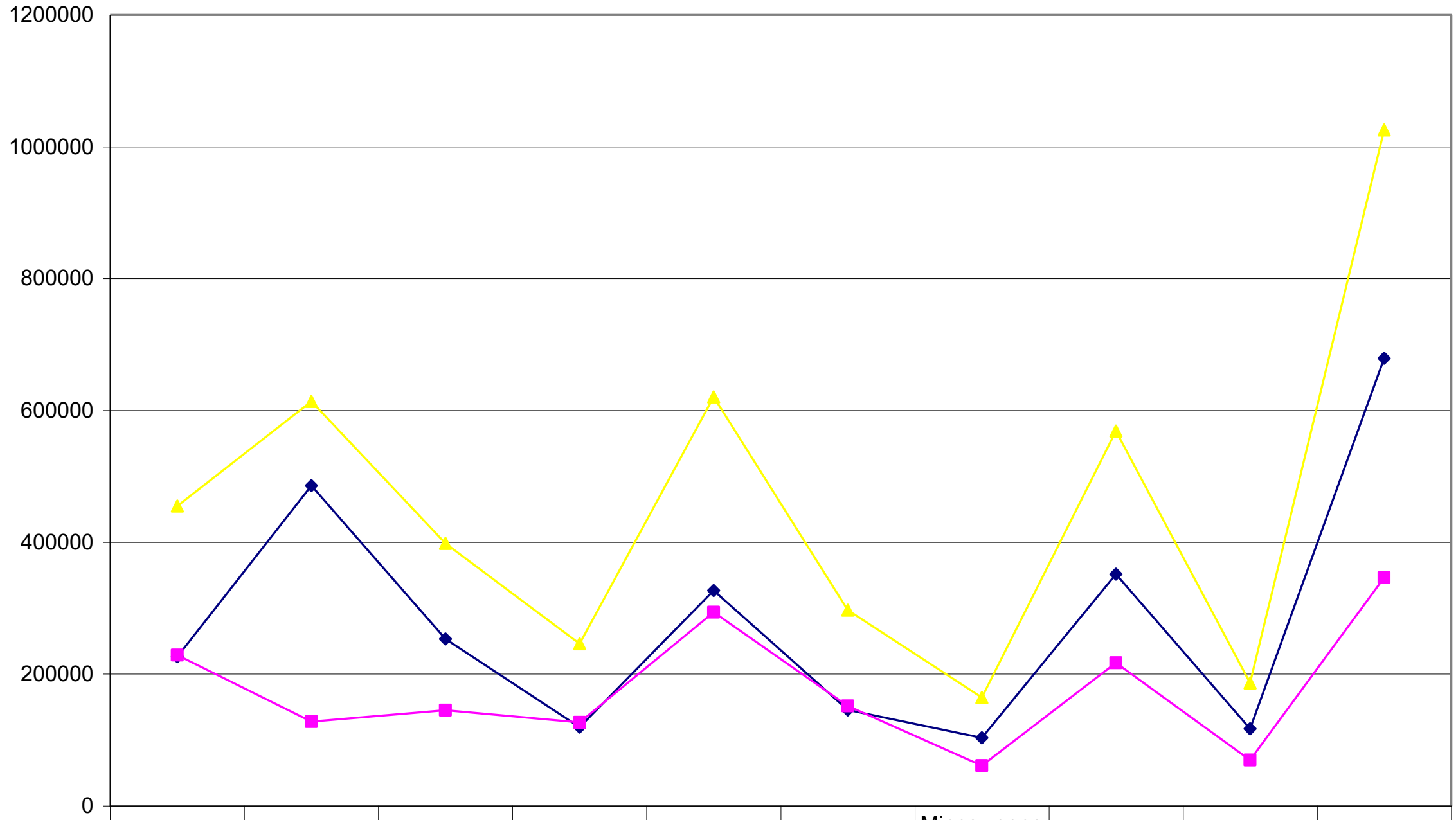

| | Arcanoid | Balote | Carrom | Fruity | Goal Englizi | Loteria | Minesweeper | Tarneeb | Taxi Escape | Trix |
|---|---|---|---|---|---|---|---|---|---|---|
| Vorg | 226233 | 485792 | 253241 | 119388 | 326755 | 145534 | 103333 | 351508 | 116884 | 679118 |
| Vr | 228589 | 127880 | 144917 | 126623 | 293749 | 151455 | 61084 | 217145 | 69630 | 346579 |
| Vnr | 454823 | 613672 | 398158 | 246011 | 620505 | 296989 | 164417 | 568653 | 186515 | 1025697 |

Figure 5: the Vorg, Vr, and Vnr of Experimental Projects.

The results of Library Investment Ratio for experimental projects are presented in figure 6. LIR indicates to the reduction ratio in software complexity, software testing cost, and software design cost. The results show that, Fruity satisfies the highest reduction ratio; Balot has the lowest reduction ratio. These projects are ordered decreasingly Fruity, Loteria, Arcanoid, Goal Englizi, Tarneeb, Taxi Escape, Minesweeper, Carrom, Trix and Balot. The results of LIR are nearly compatible with the results of reduction volume Vr.

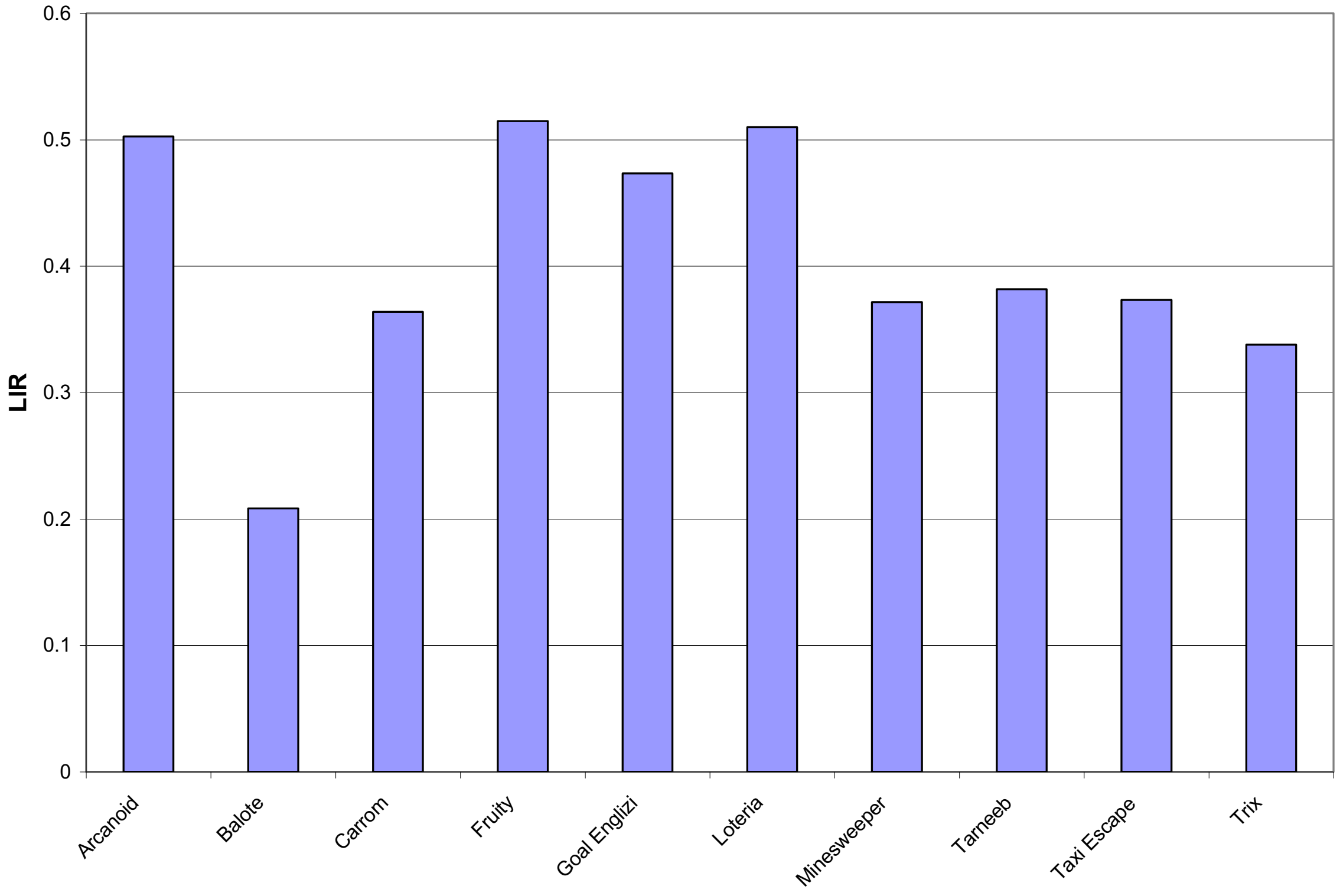

Figure 6: the Result of LIR.

Figure 7 shows the results of Library Investment Level. LIL refers to the investment level that indicates to the software productivity improvement level. The results show, that Fruity achieves the best improvement in software productivity. The projects can be ordered decreasingly based on the LIL, in which Fruity is the first one and Trix the last one. Others are Loteria, Arcanoid, Goal Englizi, Tarneeb, Taxi Escape, Minesweeper, Carrom, and Trix in descending order respectively. The results of LIL are closely compatible with the result of LIR.

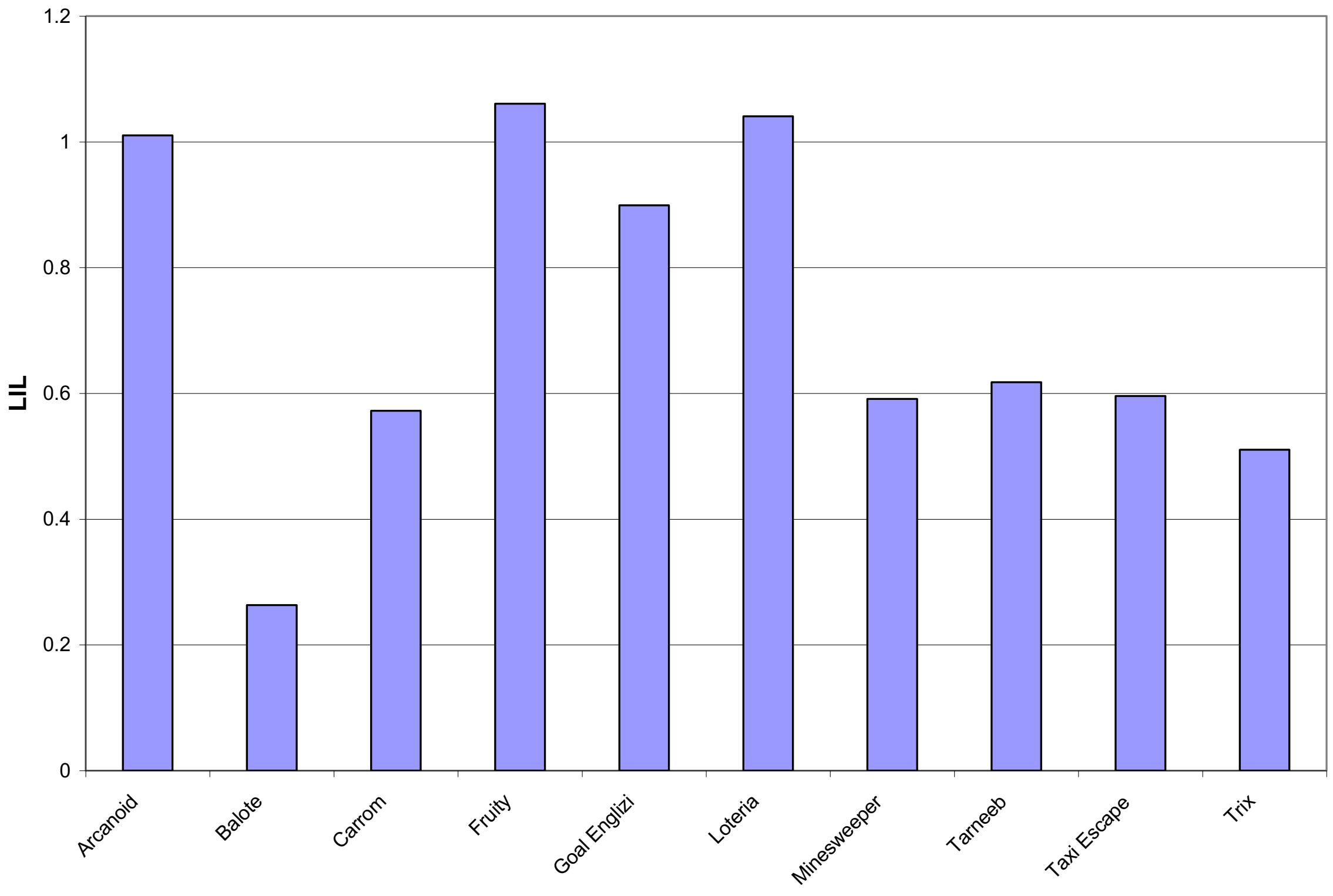

Figure 7: the Result of LIL.

The results of Program Simplicity are presented in figure 8. The results show that the PS value that is resulted from library investment is accepted because PS is closely related to Library Investment Level. Therefore, the results of PS are compatible with LIL and LIR results. The results show that Fruity has the greater PS value. Balot project has the least PS value; others are ordered decreasingly from Loteria, Arcanoid, Goal Englizi, Tarneeb, Taxi Escape, Minesweeper, Carrom, and Trix respectively.

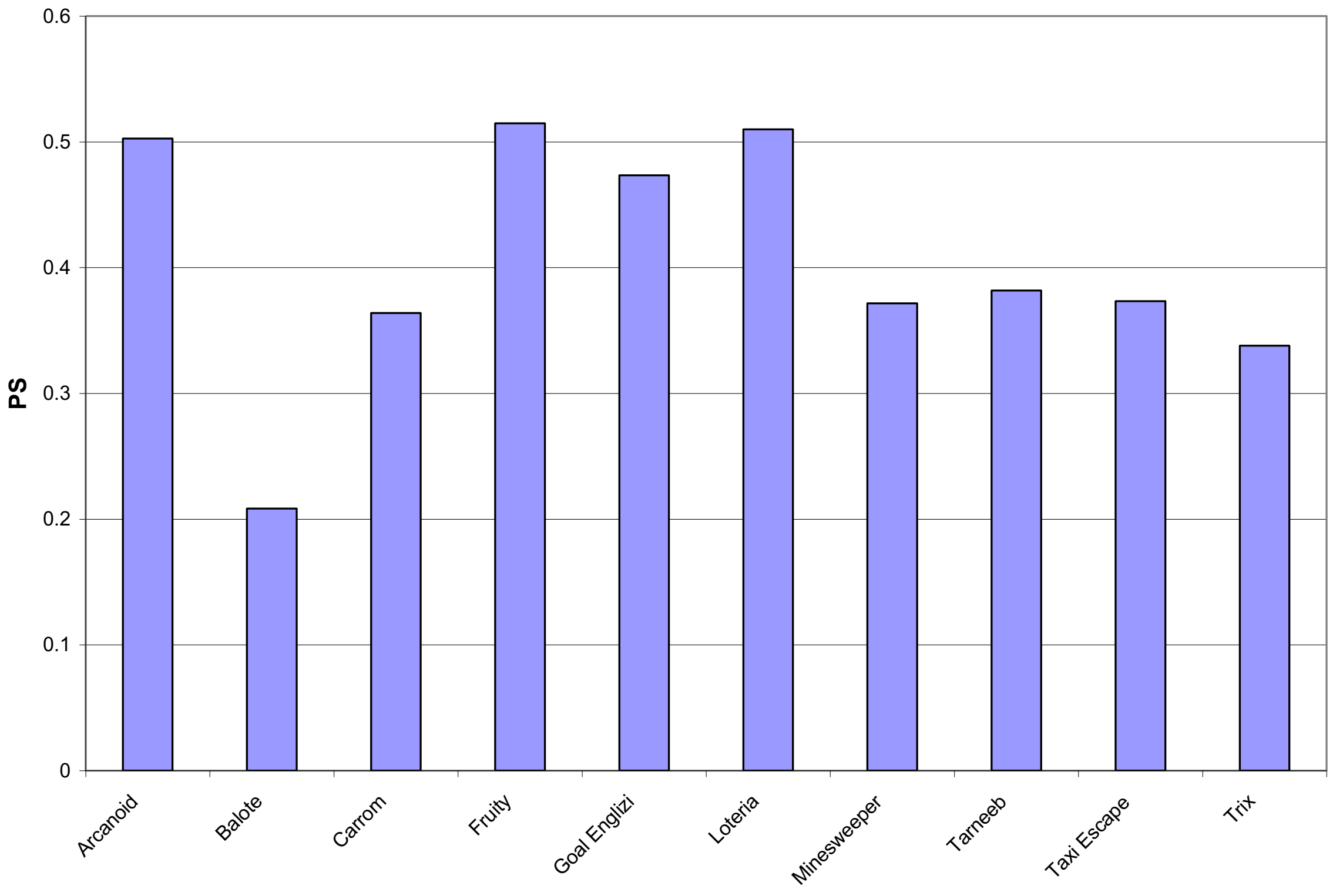

Figure 8: the Result of PS.

The comparison between RP and LIR is presented in figure 9. The results show that RP is always satisfied higher percentage than LIR, and the expected Balot value. The large gab between them is generated from the differences in the calculation methods. RP finds the reuse ratio based on the numbers of line code, but LIR deepens on the line of code by taking the content of the line of code in its consideration. Therefore, the results of the model metrics are better than Reuse Percent. For example, assume that such a line of code contains (int x =10, y=100, z = 1000), RP takes this line as one segment but LIR deepens in the line and decomposes it into small segments, which led to more accuracy when calculating the LIR metric.

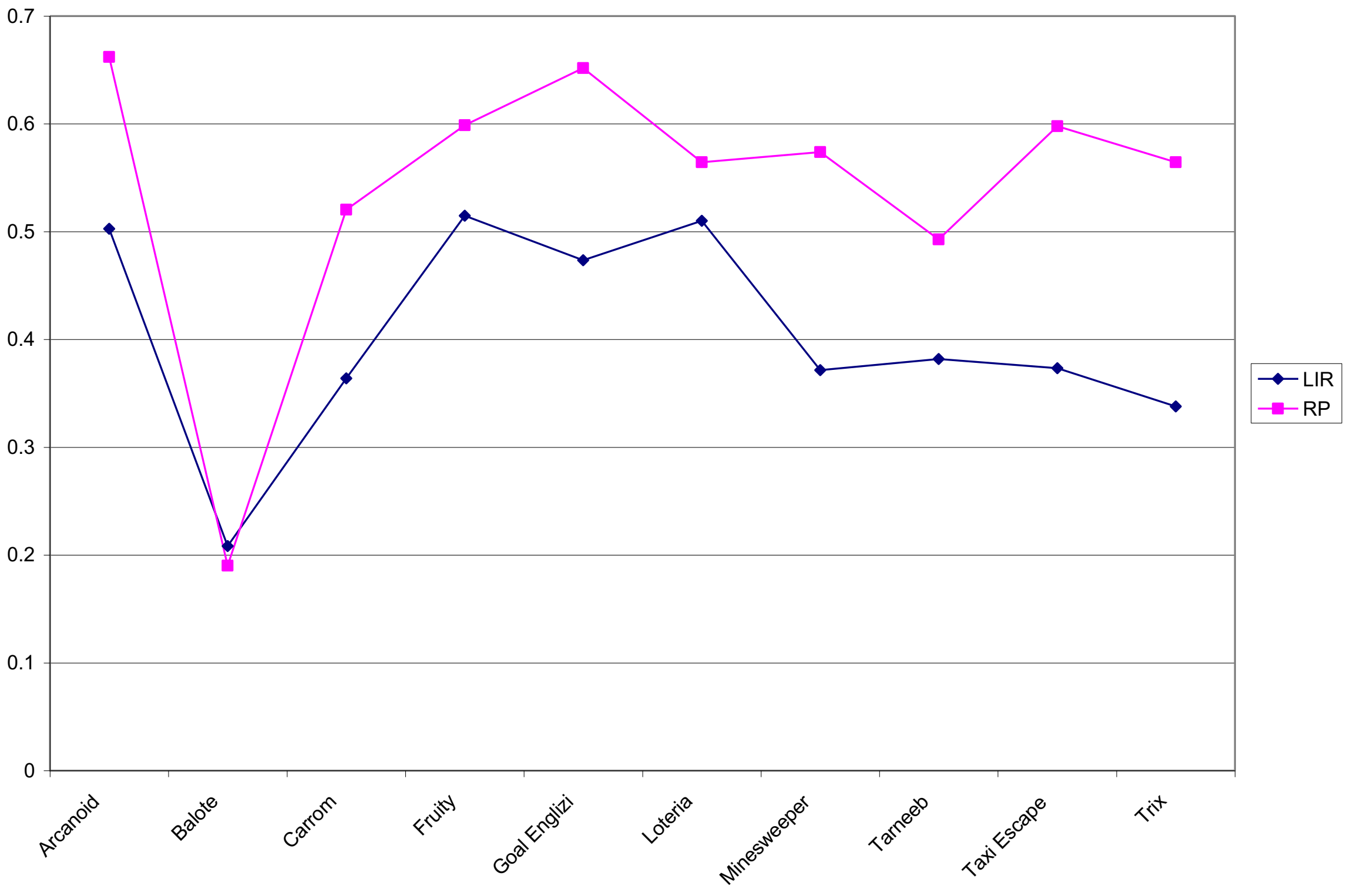

Figure 9: the Results of LIR and RP.

# Chapter Five: Conclusions and Future Works

Software reuse is very important aspect in software engineering. It improves software quality and reduces development cost. Therefore, there is a need to develop and implement software reuse metrics to assess the reuse process to make sure that the reuse process is in the correct way.

## 1. Conclusion

In this research, three library reuse metrics developed that measure library investment based on Halstead Program Volume. These metrics are Library Investment Ratio (LIR), Library Investment Level (LIL), and Program Simplicity (PS).

LIR is used to measure the reduction in software complexity, software design and testing cost that are resulted from library reuse instead of using generic programming. The formula of LIR is:

LIR = Vr / Vnr.

Where:

Vnr: program volume without reuse.

Vr: the reduction volume that is resulted from library reuse.

LIL measures the investment level, which is related to the software productivity. LIL is used as a factor that helps the decision maker to manage the available resources to improve its productivity. LIL is calculated as follow:

LIL = Vr / Vorg.

Where:

Vorg: program volume for original program.

Vr: the reduction volume that is resulted from library reuse.

Program Simplicity is used as indicator to the simplicity ratio that is resulted. PS is:

PS = 1 – (Vorg / Vnr).

Where:

Vnr: program volume without reuse.

Vorg: program volume for original program.

The model is implemented using Java programming language. The model is applied into several projects that are collected from Maysalward Inc Company to find the results of the developed model. The results show that the library reuse improves software quality, and productivity. It reduces production time, and development cost.

The results of the model are compared with Reuse Percent (RP). The results show that the model introduces better results than RP. Because the model is deepening in the source code more than the RP does.

## 2. Future Directions

There are many future Directions that are concluded during this research. These are closely related to software quality. The following list includes a few of future works that researchers can pursue:

1- Inheritance and Polymorphism are very important aspects in object oriented programming that are used to reduced software complexity. Thus, there is a need to develop new success software metrics that measure Inheritance investment, and Polymorphism investment.

2- An empirical analyzes should be applied to find the relationship between the model metrics and other software quality attributes. Such as customer satisfaction, software understandability, maintainability, and software readability.

# Arabic Abstract

## مخطط جديد لقياس استثمار المكتبات البرمجية

إعداد: أنس محمد حسن شطناوي

## الملخص

تعتبر صناعة البرمجيات احدى اهم الصناعات الاستراتيجية في وقتنا الحاضر, حيث ارتبطت ارتباطا وثيقا بحياة المجتمعات, لذلك ازدهر علم هندسة البرمجيات بشكل مميز لدعم هذه الصناعة. تعتبر جودة البرمجيات واحدة من اهم التحديات في علم هندسة البرمجيات, و لجودة البرمجيات ابعاد عدة, تختلف باختلاف وجهة نظر المستخدم واحتياجاته. وبالتالي قد تقودنا هذه الابعاد الى صعوبة في قياس وتعريف ماهية جودة البرمجيات بشكل كافي. لذا تعتبر عملية قياس الجودة واحده من اهم المجالات البحثية التي تواجه الباحثين في هذا المجال.

ان استخدام المكتبات البرمجية بدلا من الطرق البرمجية التقليدية يساعد في زيادة جودة المنتج البرمجي, حيث ان هذه المكتبات قد اعدت وفحصت مسبقا. فاستخدام هذه المكتبات يساهم في تقليل الكلفة اللازمة في مراحل عملية الانتاج.

ان الهدف من هذه الاطروحة هو اقتراح نموذج قياس جديد يهدف الى قياس الكلف الموفَّرة جراء استخدام المكتبات البرمجية بدلا من الطرق البرمجية التقليدية. هذا النموذج يحتوي على ثلاثة مقاييس جديده هي مقياس نسبة استثمار المكتبات البرمجة (LIR) و مقياس درجة استثمار المكتبات البرمجة (LIL) و مقياس درجة السهولة الناتجة عن استخدام المكتبات البرمجية (PS).

تم دراسة المقاييس تجريبا ومقارنتها مع نسبة اعادة الاستخدام (Reuse Percent) وذلك من خلال تطبيقها على عشرة مشاريع برمجية. وقد اظهرت النتائج ان المقاييس المقترحة تعطي نتائج ادق من مقياس نسبة اعادة الاستخدام.